\documentclass[12pt,preprint]{aastex}
\usepackage{graphicx}
\title{Discrete and Continuous Ejection Models of the Radio Source Associated with GW170817}
\author{Brian Punsly\altaffilmark{1}}
 \altaffiltext{1}{1415 Granvia Altamira, Palos Verdes Estates CA, USA
90274: ICRANet, Piazza della Repubblica 10 Pescara 65100, Italy and
ICRA, Physics Department, University La Sapienza, Roma, Italy,
brian.punsly@cox.net}
\begin{document}
\begin{abstract} The gravity wave source, GW170817, and associated gamma ray burst (GRB), GRB 170817A, produced radio emission that was detected in multiple epochs of Very Long Baseline Interferometry (VLBI) and with broadband radio photometry. Three unique pieces of observational evidence were determined: a discrete radio emitting region that moves with an apparent velocity of $\approx 4$c, the discrete region includes all of the radio flux, and there is likely a synchrotron self absorption (SSA) spectral turnover on day $\sim 110$ and day $\sim 160$ after ejection. This unprecedented wealth of data for a GRB provides a unique opportunity to understand the radio emitting plasma that was ejected by the putative merger event. The velocity can constrain the kinematics and the SSA turnover has been used to constrain the size to much smaller than can be done with an unresolved VLBI image, allowing one to estimate the associated plasmoid size directly from the data and improve estimates of the energetics. Models of the radio emission for both a turbulent, protonic, discrete ballistic ejection and a high dissipation region within an otherwise invisible Poynting flux dominated positron-electron jet are considered. On days $\sim 110$ and $\sim 160$ post-merger, for the range of models presented, the jet power is $2\times 10^{39} - 8\times 10^{40} \rm{ergs/s}$  and the ballistic plasmoid kinetic energy is $3\times 10^{45} - 1.5\times 10^{47} \rm{ergs}$. Even though only valid after day 110, this independent analysis augments traditional GRB light curve studies, providing additional constraints on the merger event.
\end{abstract}

\section{Introduction}
The gravitational wave (GW) detection, GW170817, by LIGO and VIRGO on August 17 2017 was accompanied by a short gamma ray burst (GRB), GRB 170817A, detection by FERMI \citep{abb17,abb18,gol17}. The GRB is believed to be associated with a merger event with a stellar mass compact remnant \citep{abb18,rue18}. This is the first association of an electromagnetic signal with a GW and has received extensive monitoring over the last year. Monitoring has revealed a wealth of information that has generally not been available for other GRBs. There have been many attempts to analyze this event by methods traditionally used for other GRBs with less observational information, namely a light curve powerlaw analysis \citep{hal17,moo18,moo19,tro18,tro19,dob18}. The results suggested a more complicated time evolution than was typical for GRBs in order to explain the rising X-ray emission over the first three months. The detection of a moving radio component with an apparent velocity, $v_{\rm{app}}\approx 4 \rm{c}$ is direct observational evidence that restricts the kinematics \citep{moo19}. The discrete component that was detected with Very Long Baseline Interferometry, VLBI, ($\sim 10,000$ km baselines) at 4.5 GHz on day 75 - day 230 after the merger had the same flux density as that found on $\sim 25$ km baselines with the Jansky Very Large Array, JVLA \citep{moo19,ghi18}. This indicates
 that all of the detected radio flux from this event is contained in the single moving unresolved component. This appears to be directly analogous to the most powerful ejections from stellar mass compact objects in the Galaxy, the superluminal discrete ejections or major flares \citep{mir94,fen99}. Thus motivated, a treatment of the radio emitting ejection associated with GW170817 in direct analogy to previous analysis of stellar mass black holes with superluminal discrete ejections in the Galaxy is presented.
\par  In the Galaxy, discrete superluminal ejections have been interpreted as either a moving, strong dissipation region in a continuous relativistic jet or a discrete ballistic ejection \citep{fen04,pun12}. Similarly, the model chosen for GW170817 is one in which all of the emission is from a single moving unresolved component with no detected emission from a stationary core or a smooth background jet. Since the emission is unresolved, no structure is ascertained and a simple spherical homogeneous volume is assumed. The region can be considered a discrete ballistic plasmoid that was ejected along a preferred axis (such as a total angular momentum axis) as a byproduct of the merger event. Alternatively, if one assume a continuous jet then the spherical dissipation region might be the advancing head of the jet, akin to a radio lobe or hot-spot in an extragalactic radio source. The unresolved VLBI image restricts the size to $< 10^{17}$ cm, \citep{ghi18}. However, for the spherical volume, if the spectrum appears as a powerlaw with a low frequency (synchrotron self-absorbed, SAA) turnover, one can obtain an estimate of the spatial dimension that can be much less than the unresolved image \citep{rey09}. This dimensional estimate can be used to improve estimates of the energetics of the ejection if the bulk flow Doppler factor of the plasmoid, $\delta$, is estimated from the observed kinematics \citep{pun12,lig75},
\begin{equation}
\delta = \frac{\gamma^{-1}}{1-\beta \cos{\theta}},\; \gamma^{-1} = 1- \beta^{2}\;,
\end{equation}
where $\beta$ is the normalized three-velocity of bulk motion and $\theta$ is the angle of the motion to the line of sight (LOS) to the observer. These methods are presented in Section 2. They were first developed to describe discrete ejections from the quasar, Mrk 231, and was later applied to the major ejections from the Galactic black hole GRS 1915+105 \citep{rey09, pun12}.

\begin{figure}
\begin{center}
\includegraphics[width=75 mm, angle= 0]{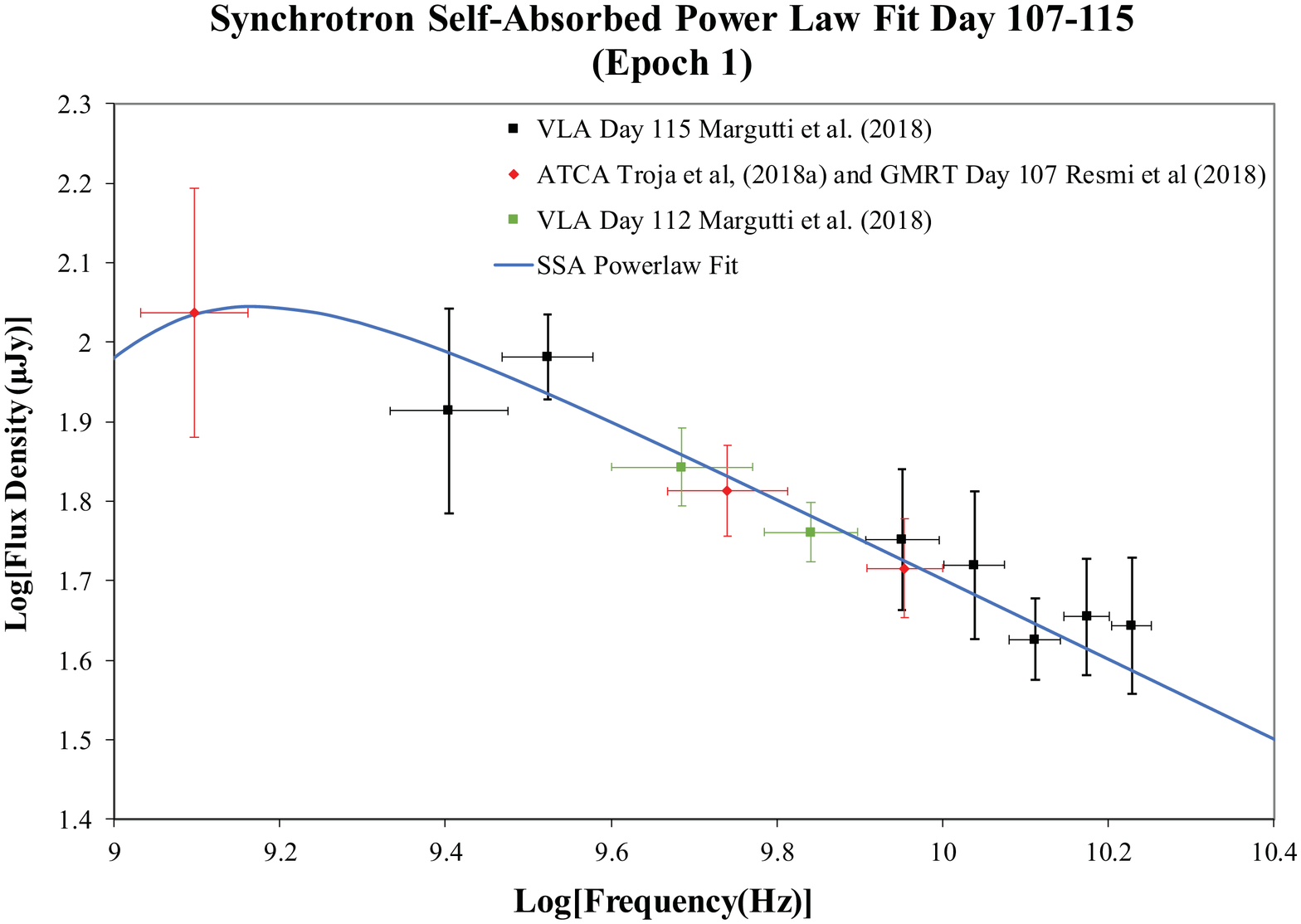}
\includegraphics[width=75 mm, angle= 0]{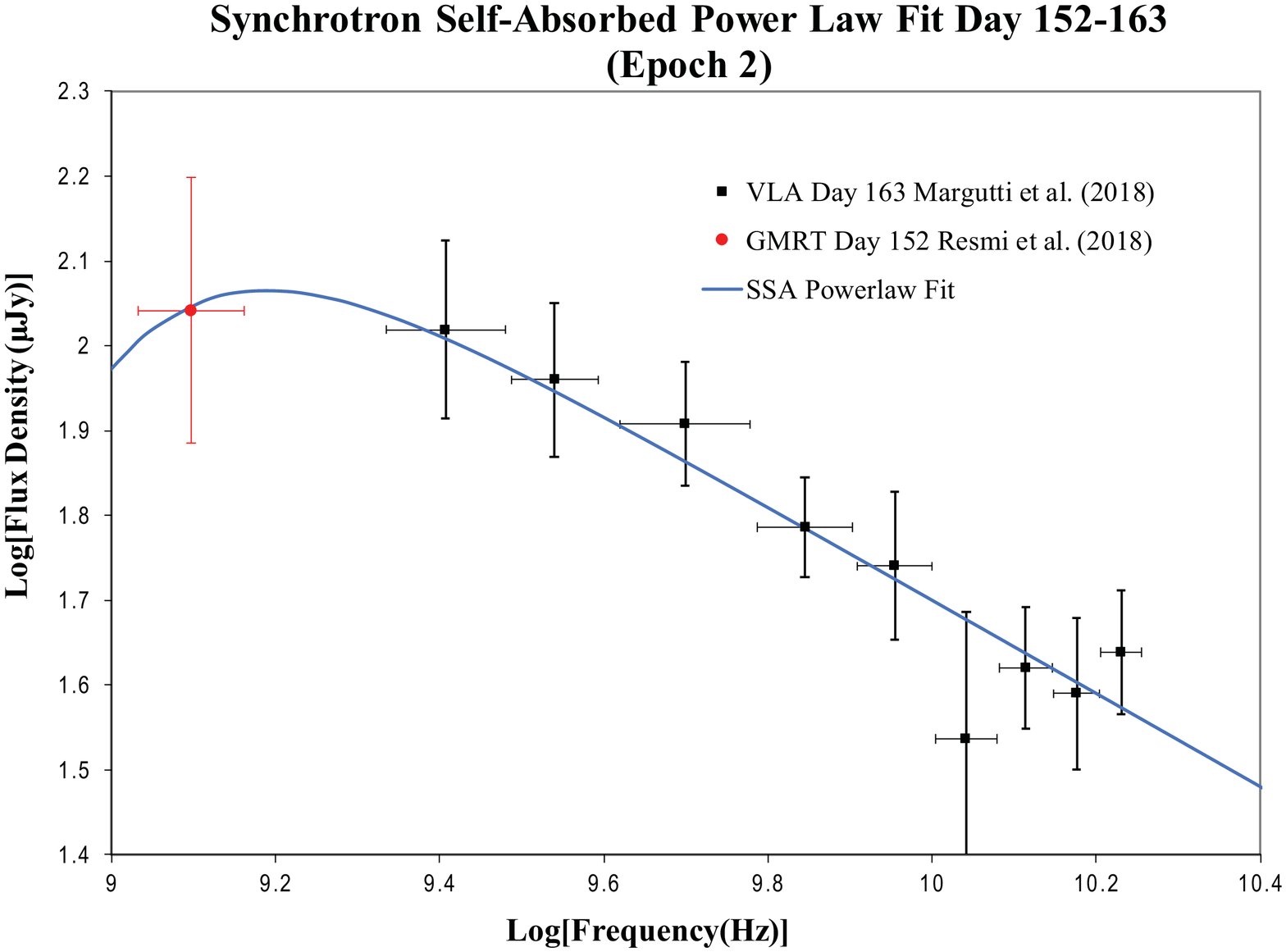}
\includegraphics[width=75 mm, angle= 0]{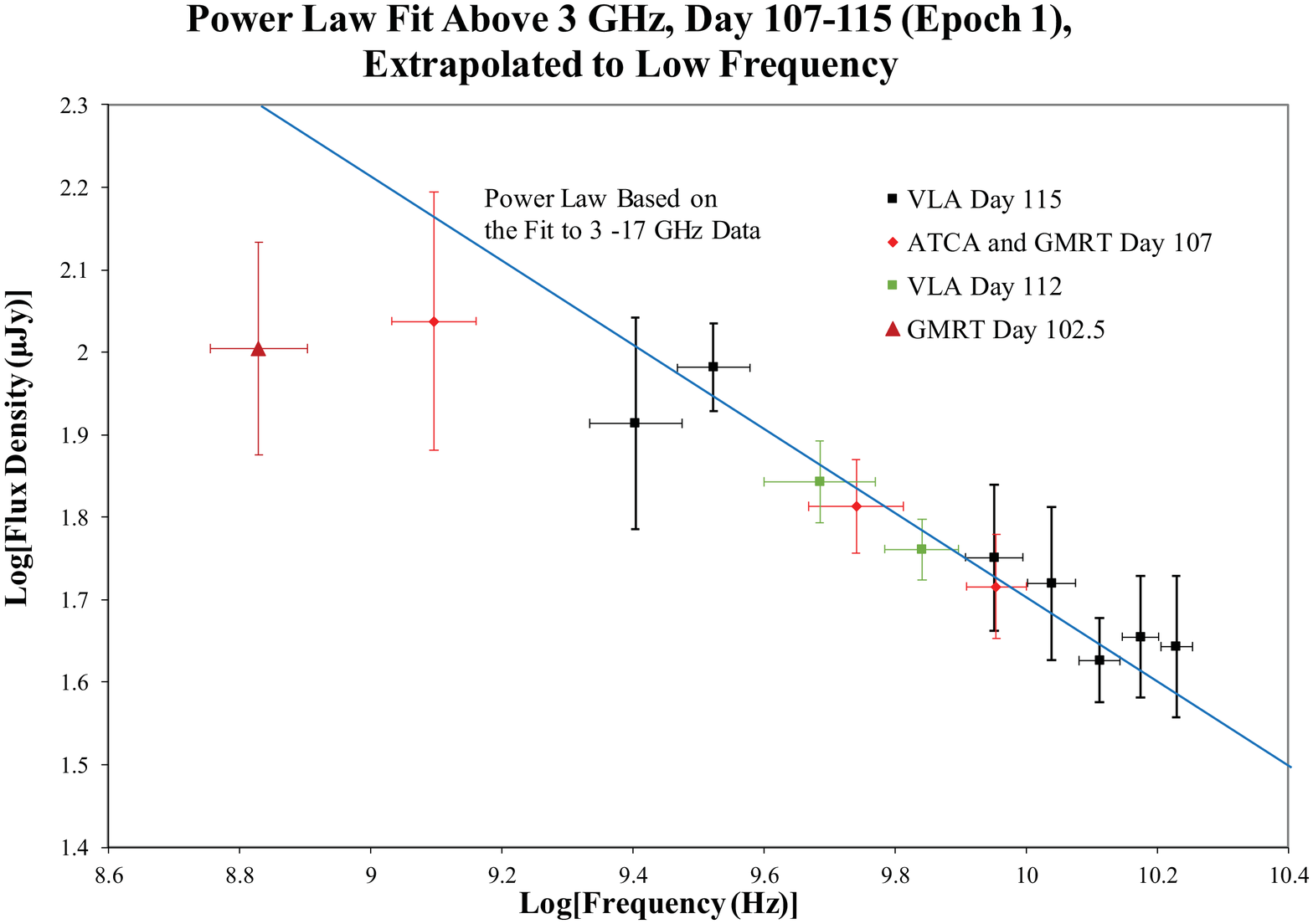}
\includegraphics[width=75 mm, angle= 0]{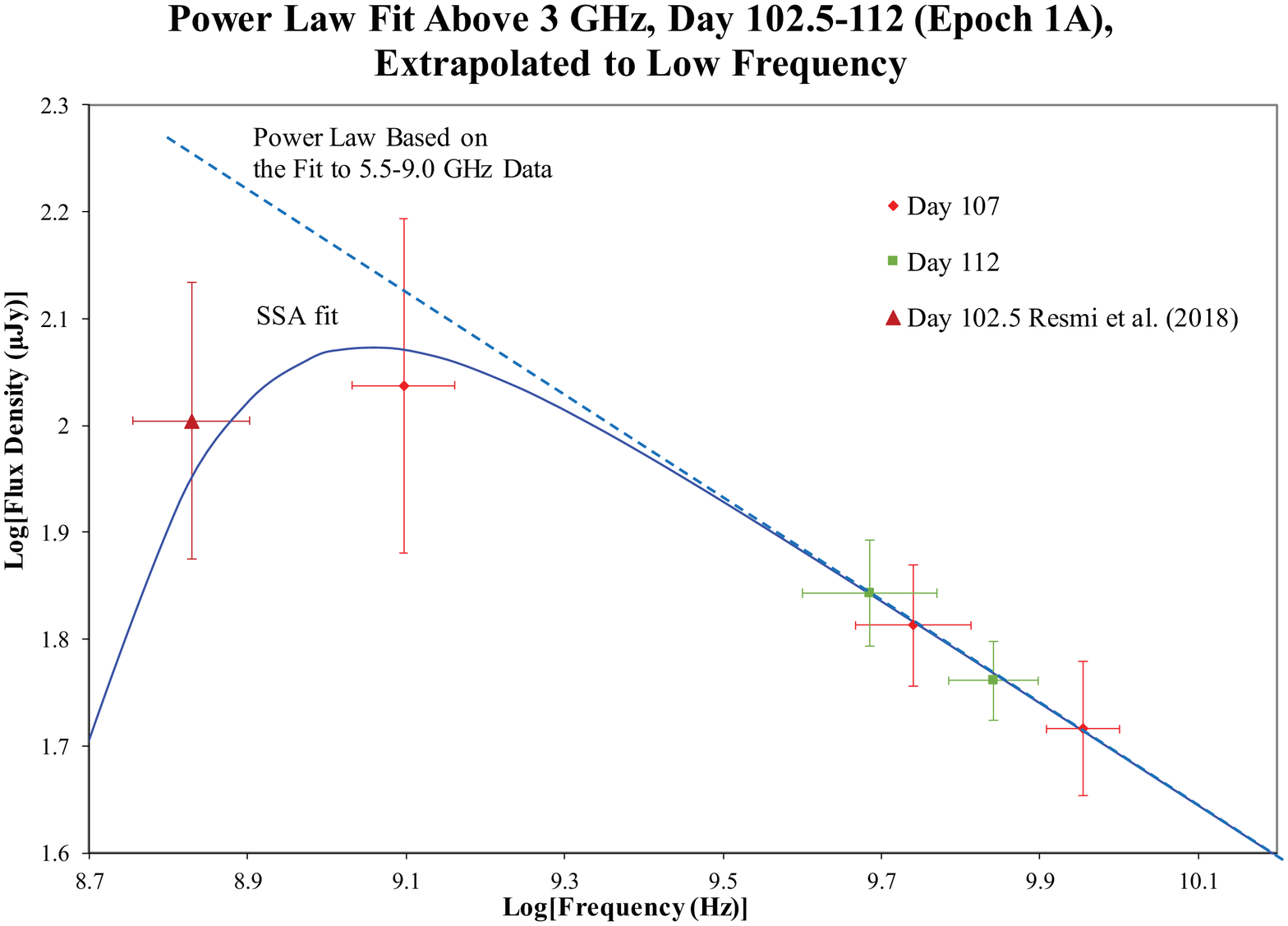}
\caption{The top left (right) frame is the radio data from epoch 1 (epoch 2). The SSA powerlaw fits of Equation (5) are the continuous curves (see Sections 2 and 3). The bottom frames extrapolate the powerlaw fit above 3 GHz to lower frequency revealing an excess at 675 MHz.}
\end{center}
\end{figure}
\par In Section 3, the model is constrained by the radio spectra on days $\sim 110$ and $\sim 160$. These constraints and the apparent velocity of 4c are used to explore the physical solution space of the models in Section 4. The results are analyzed in Section 5.
\section{Synchrotron Self-Absorbed Homogeneous Plasmoids}
Since the ejection is unresolved, the simple homogeneous spherical volume model of \citet{van66} is chosen. The strategy will be to evaluate in the rest frame then transform the results to the observer's frame for comparison with observation. The underlying powerlaw for the flux density is defined
by $S_{\nu}(\nu= \nu_{o}) = S\nu_{o}^{-\alpha}$, where $S$ is a constant. Observed quantities will
be designated with a subscript, ``o", in the following expressions.
The SSA attenuation coefficient in the plasma rest frame (noting
that the emitted frequency is designated by $\nu$) is given by \citep{gin69},
\begin{eqnarray}
&& \mu(\nu)=\overline{g(n)}\frac{e^{3}}{2\pi
m_{e}}N_{\Gamma}(m_{e}c^{2})^{2\alpha} \left(\frac{3e}{2\pi
m_{e}^{3} c^{5}}\right)^{\frac{1+2\alpha}{2}}\left(B\right)^{(1.5
+\alpha)}\left(\nu\right)^{-(2.5 + \alpha)}\;,\\
&& \overline{g(n)}= \frac{\sqrt{3\pi}}{8}\frac{\overline{\Gamma}[(3n
+ 22)/12]\overline{\Gamma}[(3n + 2)/12]\overline{\Gamma}[(n +
6)/4]}{\overline{\Gamma}[(n + 8)/4]}\;, \\
&& N=\int_{\Gamma_{min}}^{\Gamma_{max}}{N_{\Gamma}\Gamma^{-n}\,
d\Gamma}\;,\; n= 2\alpha +1 \;,
\end{eqnarray}
where $\Gamma$ is the ratio of lepton energy to rest mass energy,
$m_{e}c^2$ and $\overline{\Gamma}$ is the gamma function. $B$ is the magnitude of the total
magnetic field. The powerlaw spectral index for the flux density is
$\alpha=(n-1)/2$. The low energy cutoff, $E_{min} = \Gamma_{min}m_{e}c^2$,
is not constrained by the data. The SSA opacity in the observer's frame, $\mu(\nu_{o})$, is obtained by direct substitution of $\nu =
\nu_{o} / \delta$ into Equation (2). The homogeneous approximation yields a simplified solution to the radiative transfer equation \citep{gin65,van66}
\begin{eqnarray}
&& S_{\nu_{\o}} = \frac{S_{o}\nu_{o}^{-\alpha}}{\tau(\nu_{o})} \times \left(1 -e^{-\tau(\nu_{o})}\right)\;, \; \tau(\nu_{o}) \equiv \mu(\nu_{o}) L\;, \; \tau(\nu_{o})=\overline{\tau}\nu_{o}^{(-2.5 +\alpha)}\;,
\end{eqnarray}
where $\tau(\nu)$ is the SSA opacity, $L$ is the path length in the rest frame of the plasma, $S_{o}$ is a normalization factor and $\overline{\tau}$ is a constant. There are three unknowns in Equation (5), $\overline{\tau}$, $\alpha$ and $S_{o}$. These are three constraints on the following theoretical model that are estimated from the observational data.
\par The theoretical spectrum is parameterized by Equations (2) - (5) and the synchrotron emissivity that is given in \citet{tuc75} as
\begin{eqnarray}
&& j_{\nu} = 1.7 \times 10^{-21} [4 \pi N_{\Gamma}]a(n)B^{(1
+\alpha)}\left(\frac{4
\times 10^{6}}{\nu}\right)^{\alpha}\;,\\
&& a(n)=\frac{\left(2^{\frac{n-1}{2}}\sqrt{3}\right)
\overline{\Gamma}\left(\frac{3n-1}{12}\right)\overline{\Gamma}\left(\frac{3n+19}{12}\right)
\overline{\Gamma}\left(\frac{n+5}{4}\right)}
       {8\sqrt\pi(n+1)\overline{\Gamma}\left(\frac{n+7}{4}\right)} \;.
\end{eqnarray}
One can transform this to the observed flux density, $S(\nu_{o})$,
in the optically thin region of the spectrum using the relativistic transformation relations from
\citet{lin85},
\begin{eqnarray}
 && S(\nu_{o}) = \frac{\delta^{(3 + \alpha)}}{4\pi D_{L}^{2}}\int{j_{\nu}^{'} d V{'}}\;,
\end{eqnarray}
where $D_{L}$ is the luminosity distance and in this expression, the
primed frame is the rest frame of the plasma.

\begin{figure}
\begin{center}
\includegraphics[width=90 mm, angle= 0]{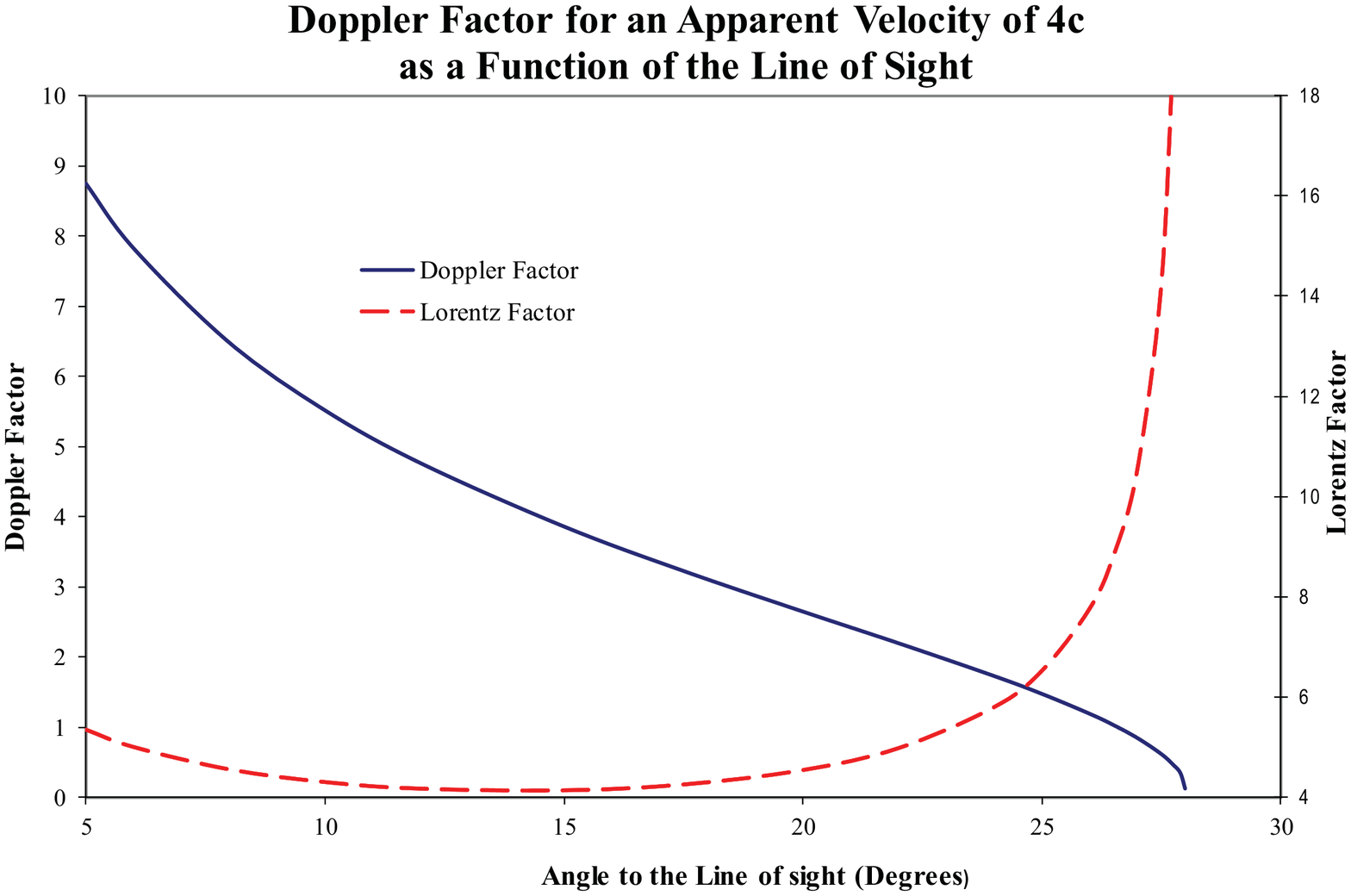}
\caption{The detection of superluminal motion, 4c, restricts $\delta$, LOS and $\gamma$.}
\end{center}
\end{figure}
\section{Fitting the Data}
The epochs of days 107 -115 (epoch 1) and days 152 - 163 (epoch 2), near the peak of radio flux light curves, are densely sampled in frequency from 1.25 - 17 GHz. Dense frequency sampling is essential for an accurate estimate of $\alpha$ since the flux density is very low, $\sim 20 -100 \mu$ Jy. ``Small" unanticipated systematic errors can drastically skew an individual measurement. This is clear by the outlier points in Figure 1. The other essential feature is the need for low frequency GMRT measurements in order to determine the magnitude of the SSA. The ejection is clearly resolved from the host galaxy nuclear radio source by GMRT at 1.25 GHz, but only marginally resolved at 675 MHz \citep{res18}. The analysis of this paper relies primarily on the modeling of epochs 1 and 2 that are not contemporaneous with the difficult 675 MHz Gaussian model fits.
\par Mathematically, the theoretical determination of $S_{\nu}$ depends on 7 parameters in Equations (2)-(8),
$N_{\Gamma}$, $B$, $R$ (the radius of the sphere), $\alpha$, $\delta$, $E_{min}$ and $E_{max}$, yet there are only 3 constraints from the observation $\overline{\tau}$, $\alpha$ and $S_{o}$, it is an under determined system of equations. Most of the particles are at low energy, so the solutions are insensitive to $E_{max}$. In order to study the solution space, $\delta$ and $E_{min}$ are pre-set to a 2-D array of trial values. For each trial value one has 4 unknowns, $ N_{\Gamma}$, $B$, $R$ and $\alpha$ and 3 constraints for each model. Thus, there is an infinite 1 dimensional set of solutions for each pre-assigned $\delta$ and $E_{min}$ that results in the same spectral output. First, a least squares powerlaw fit, with uncertainty in both variables. is made between $\sim$ 3-17 GHz \citep{ree89}. This fixes $S_{o}$ and $\alpha$ in Equation (5). An arbitrary $B$ is chosen. $N_{\Gamma}$ and $R$ are then iteratively varied to produce this fitted $S_{o}$ and a value of $\overline{\tau}$ that minimizes the least squares residuals relative to the GMRT data in Figure 1. Another value of $B$ is chosen and the process is repeated until the solution space is spanned for the pre-assigned $\delta$ and $E_{min}$. For epoch 1 (epoch 2), $\alpha = 0.49 \pm 0.05$ ($\alpha=0.54 \pm 0.06$).
\par The bottom frames of Figure 1 indicate that the powerlaw derived from the higher frequency data greatly exceeds the 675 MHz flux density, justifying the SSA model over the powerlaw model. Epoch 1A (bottom right), day 102.5-112, is an attempt at a quasi-simultaneous fit that includes the 675 MHz data. This fit might be less robust due to the difficult 675 MHz data reduction.
\begin{figure}
\begin{center}
\includegraphics[width=125 mm, angle= 0]{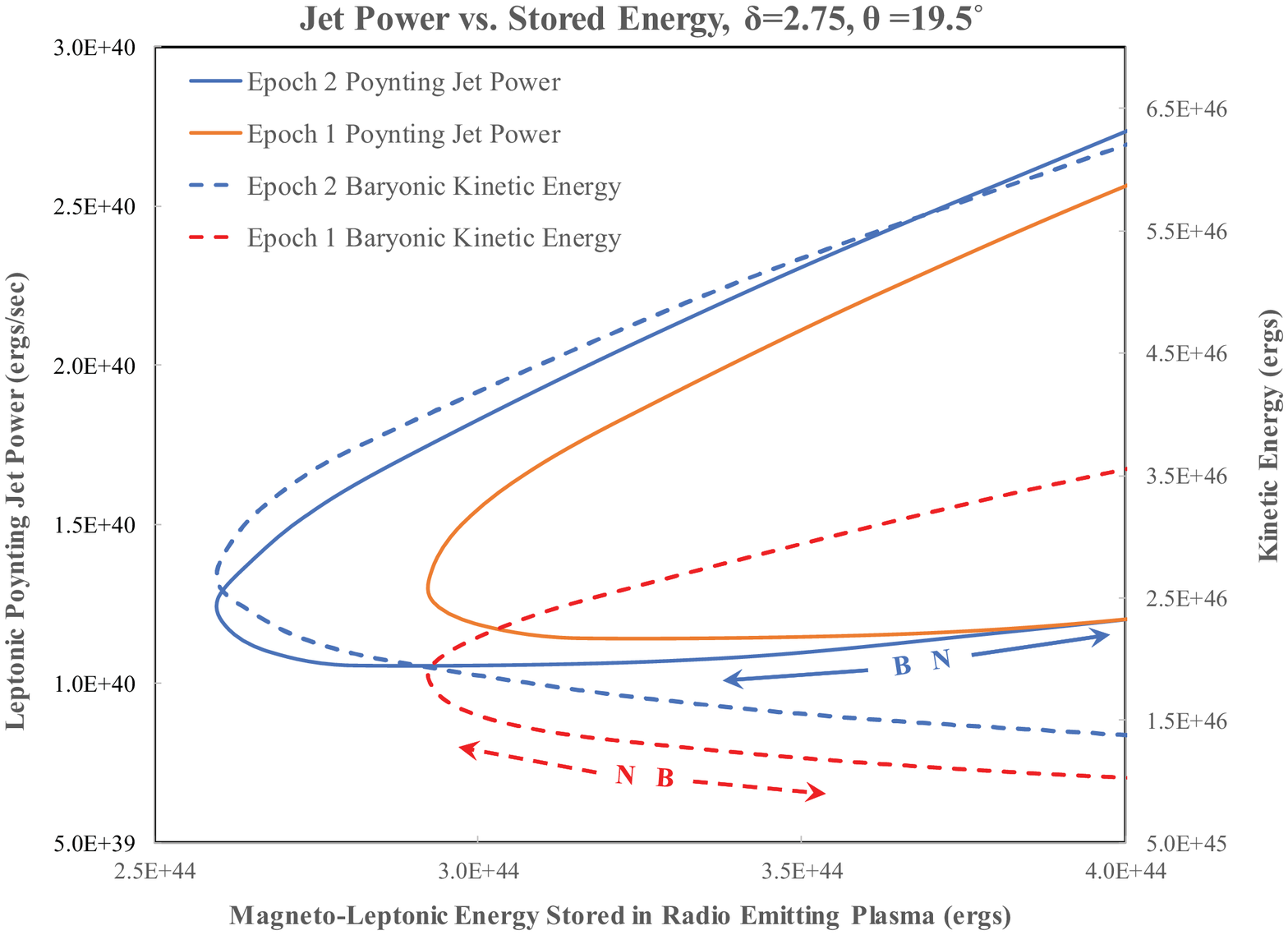}
\includegraphics[width=125 mm, angle= 0]{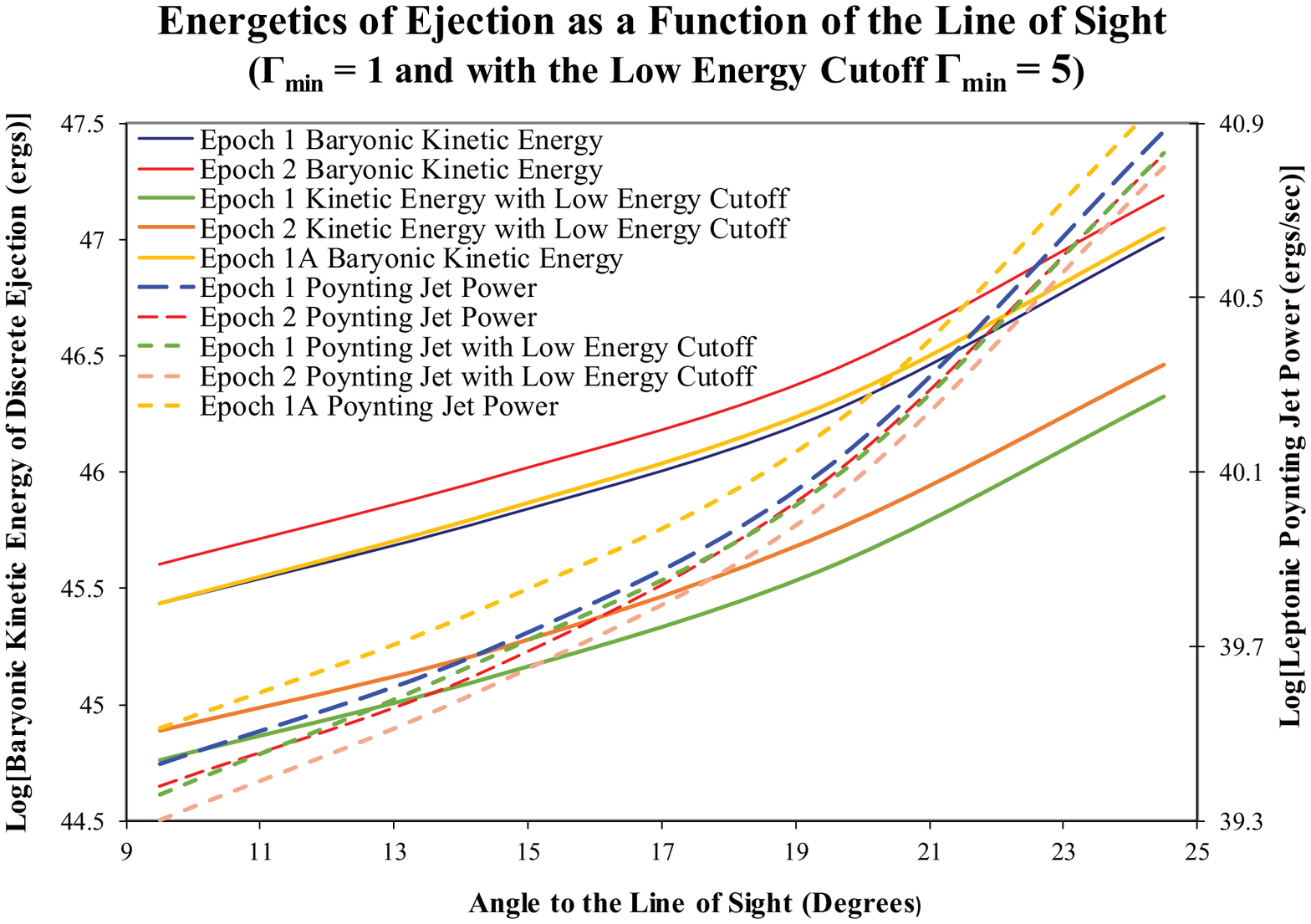}
\caption{The top panel shows the dependence of bulk kinetic energy of a discrete ejection and the jet power on $E_{lm}$ for one LOS. The behavior in similar for other LOS. The bottom panel shows how they depend on the LOS, using the restrictions from Figure 2 and assuming the minimum $E_{lm}$.}
\end{center}
\end{figure}
\section{Specific Spherical Models}
The apparent velocity constrains the kinematics of the plasmoid \citep{ree66}:
\begin{equation}
v_{app}/c=\beta_{app} = \beta \sin{\theta}/(1- \beta \cos{\theta})\approx 4\;.
\end{equation}
Equations (1) and (9) are combined in Figure 2 in order to restrict $\delta$. The superluminal motion indicates a LOS $<28^{\circ}$. The top and middle panels of Figure 1 shows the fits to the data for 1-D sets of models that exist for each chosen pair of values of $\Gamma_{min}$ and $\delta$. Using Figure 2 to restrict $\delta$, the physical parameters associated with a continuous range of models are displayed graphically in Figures 3 and 4 for $\Gamma_{min}=1$ and $\Gamma_{min}=5$. For the ballistic ejection, the magnetic field is turbulent and for a jet it is mainly organized toroidal magnetic field, $B_{\phi}^{'}$, in the rest frame of the plasma. The latter is a consequence of the perfect magnetohydrodynamic (MHD) assumption and approximate angular momentum conservation in the jet \citep{bla79}. The total angular momentum flux along the jet axis in the observer's coordinate system is \citep{pun08}
\begin{equation}
L \approx k\mu r_{\perp}\gamma v^{\phi} + \frac{c}{4\pi}B^{\phi}r_{\perp}\;, \; k \equiv \frac{\mathcal{N}\gamma v^{P}}{B^{P}};,
\end{equation}
where $k$ is the perfect MHD conserved mass flux per unit poloidal magnetic flux, $B^{P}$ and $v^{P}$ are the poloidal magnetic field and velocity, $\mathcal{N}$ is the number density in the plasma rest frame, $\mu$ is the specific enthalpy and $r_{\perp}$ is the cylindrical radius. As the jet propagates, it expands and $r_{\perp}$ is much larger than at the jet base. By Equation (10), if $L$ is approximately constant and the jet is Poynting flux dominated or the mechanical angular momentum and the electromagnetic angular momentum are comparable (which can occur at the head of the jet), with large $r_{\perp}$ ,
\begin{equation}
 v^{\phi}< \frac{L}{k\mu r_{\perp}\gamma}\quad  B^{\phi}\sim r_{\perp}^{-\eta}\;, \eta \approx 1\; .
\end{equation}
By poloidal magnetic flux conservation $B^{P} \sim  r_{\perp}^{-2}$, so  $B^{\phi} \gg B^{P}$ at large $r_{\perp}$. To estimate the poloidal Poynting flux, $S^{P}$, in the plasmoid, first transform fields to the observer's frame
\begin{equation}
B^{\phi} =\gamma B_{\phi}^{'} \quad E^{\perp} = \frac{v^{P}}{c} \gamma B_{\phi}^{'} - \frac{v^{\phi}}{c} \gamma B^{P} \approx \frac{v^{P}}{c} \gamma B_{\phi}^{'} \;,
\end{equation}
where $ E^{\perp}$ is the poloidal electric field orthogonal to the magnetic field direction. The poloidal Poynting flux in the observer's frame, $S^{P}$, along the jet direction is \citep{pun08}:
\begin{equation}
S^{P}= \frac{c}{4\pi}E^{\perp} B^{\phi}  \approx \frac{c}{4\pi}\gamma^{2} \beta [B_{\phi}^{'}]^{2} \;,\; B_{\phi}^{'} \approx B\;.
\end{equation}
\par The energy content is separated into two pieces. The first is the kinetic energy of the protons, $E(\mathrm{proton})$. The other piece is named the lepto-magnetic
energy, $E(\mathrm{lm})$, and is composed of the volume integral of the leptonic internal energy density, $U_{e}$, and the magnetic field energy density, $U_{B}$. It is straightforward to compute the lepto-magnetic energy in a spherical volume,
\begin{eqnarray}
 && E(\mathrm{lm}) = \int{(U_{B}+ U_{e})}\, dV = \frac{4}{3}\pi R^{3}\left[\frac{B^{2}}{8\pi}
+ \int_{\Gamma_{min}}^{\Gamma_{max}}(m_{e}c^{2})(N_{\Gamma}E^{-n + 1})\, d\,E \right]\;.
\end{eqnarray}
Based on superluminal ejections in the Galaxy, the jet is chosen to be leptonic, but see the Conclusion \citep{fen99,pun12}. The protonic kinetic energy is
\begin{eqnarray}
 && E(\mathrm{protonic}) = (\gamma - 1)Mc^{2}\;,
\end{eqnarray}
where $M$ is the mass of the plasmoid.
\par Figures 3 and 4 show the details of the models of Section 2 that are constrained by the fits in Figure 1 and the kinematics in Figure 2. The top left frame of Figure 3 shows the dependence on $E(\mathrm{lm})$ of both the jet power (the combination of Poynting flux and bulk leptonic internal energy flux) and the baryonic bulk kinetic energy, $E(\mathrm{protonic}$), for one particular LOS. Other LOS produce similar curves. The figure shows the direction (the arrows) of change in jet power and $E(\mathrm{protonic}$) as $B$ and $\mathcal{N}$ vary. The solution with minimum $E(\mathrm{lm})$ has historically been of much interest in astrophysics, the minimum energy solution. Discrete plasmoids ejected from the Galactic black hole, GRS~1915+105, evolve toward the minimum energy configuration at late times \citep{pun12}. For a jet model, terminating hot-spots in extragalactic jets might be analogous to the high surface brightness feature detected in \citet{moo19} and it has been argued that hot-spots tend to be near minimum energy \citep{har04}. Thus motivated, the bottom panel of Figure 3 show how the jet power and $E(\mathrm{protonic}$) of the plasmoid of the minimum energy solutions (that create the fits in Figure 1) vary with the LOS. Figure 4, shows the physical parameters in the head of the jet or discrete plasmoid for the minimum energy solution, $B$, $R$ and the total number of particles (leptons for the jet and baryons for the discrete plasmoid). Note that the fit to the 675 MHz data in epoch 1A has similar jet power and particle number to epoch 1, but $R$ is 37\% larger.

\begin{figure}
\begin{center}
\includegraphics[width=125 mm, angle= 0]{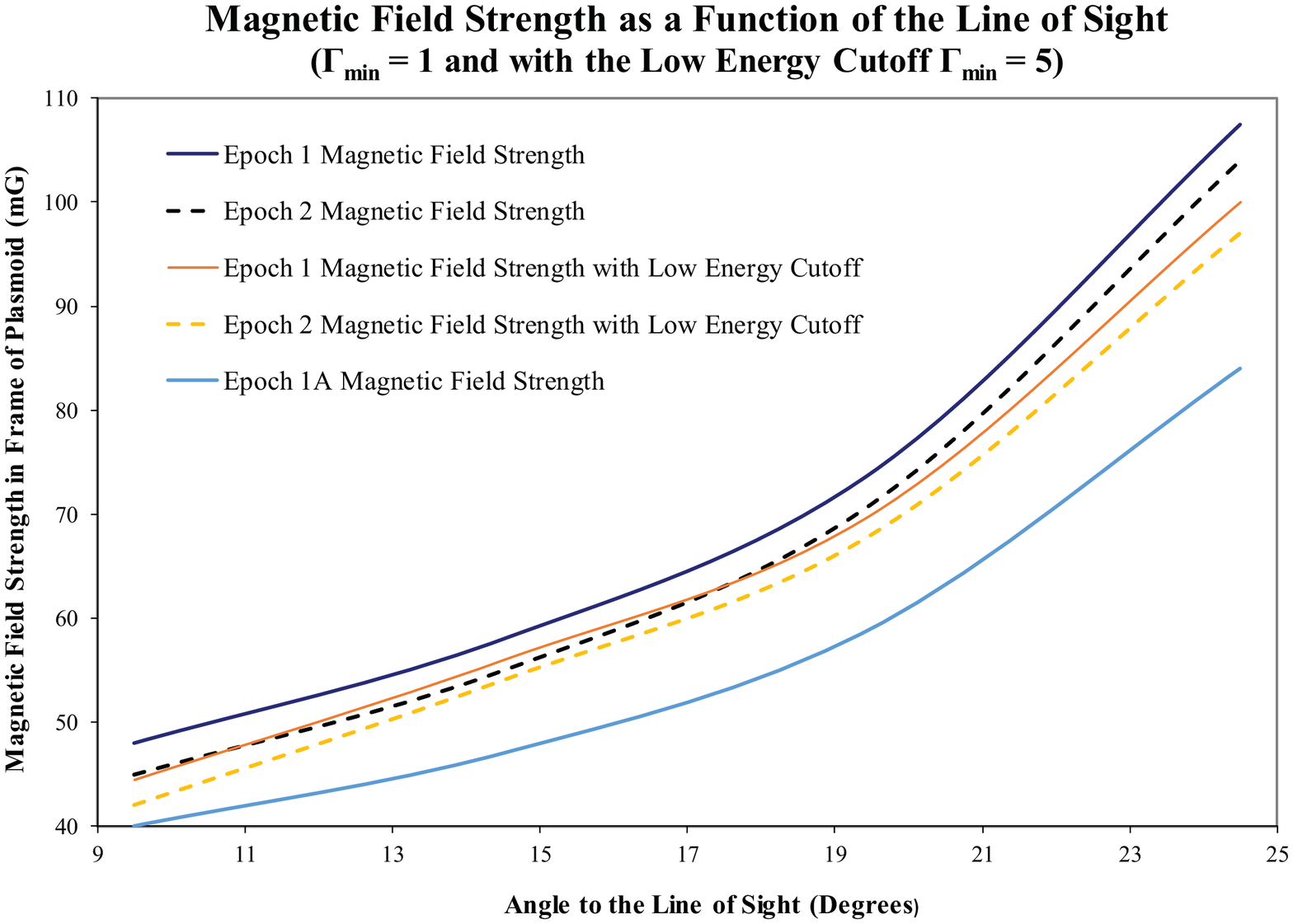}
\includegraphics[width=125 mm, angle= 0]{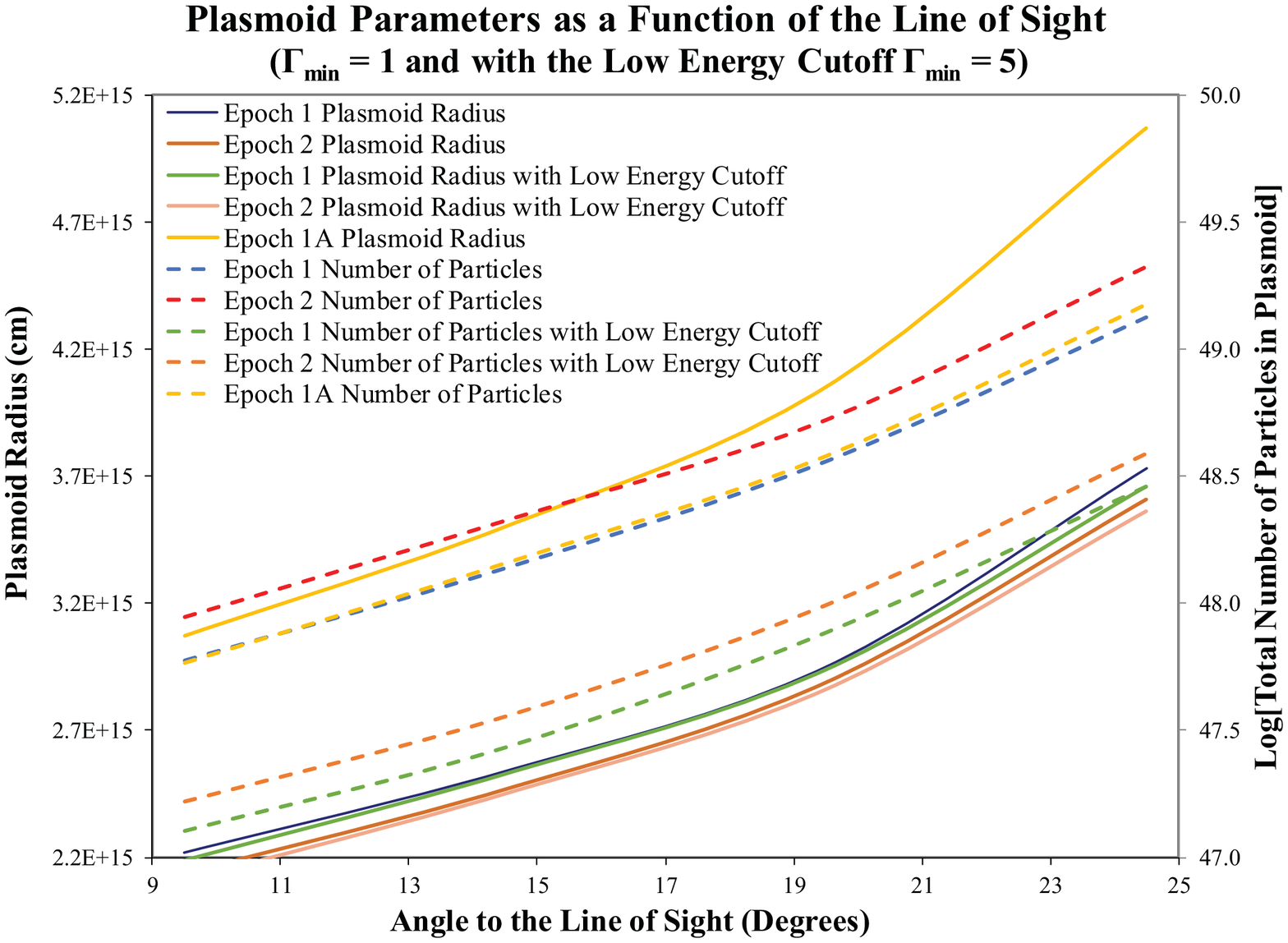}
\caption{The top (bottom) panel shows how $B$ ($R$ and the total number of particles in the plasmoid) in the spherical models depends on the LOS, assuming the minimum $E_{lm}$.}
\end{center}
\end{figure}
\section{Analysis of Results}
Due to the low value of the flux densities used to generate these results, both the epoch 1 and epoch 2 data are valuable for constraining the parameters of the outflow. We note from Figures 3 and 4 that $R$, $B$ and Poynting flux are very similar in both epochs and these are considered the most robust results. This section is primarily concerned with constraints imposed by X-ray losses and the temporal behavior of the flare decay.
\par Temporal decay is a standard tool in the analysis of GRBs. This analysis is hindered by the low flux densities. For example, we do not use the ATCA 7.25 GHz data during the decay after day 183 because in spite of long $\sim 11 $ hour observations, the flux calibration has a potential 25\% systematic error in some or all of the observations \citep{moo20,tro19}. We use only the 3 GHz flux densities form the JVLA in top panel of Figure 5, they are larger and the JVLA has higher sensitivity \citep{moo20,ale18}. The dashed yellow line is the least squares fit with uncertainty in both variables, the powerlaw decay constant, $S_{\nu}(t)\sim t^{-\zeta}$, is $\zeta =2.30\pm0.24$. First compare this to the discrete uniform, spherical, adiabatic model of \citet{van66} that begins expanding into a uniform medium after day 180. The radius of the plasmoid scales like $R(t)\sim t^{\omega}$, with $S_{\nu}(t)\sim t^{-2n\omega}$, $\omega = 0.4$ \citep{van66}. The analysis raises the uncertainty imposed by small flux densities, \citet{ale18} find $\alpha =0.74\pm0.2$ or $n=2.48$ on day 217, yet \citet{moo20} estimate $\alpha =0.584 $ or $n=2.17$ from their analysis. A value of $\omega = 0.42$ and $n=2.48$ is implemented in Figure 5 which results in $\zeta =2.08$ within the standard error of the least squares fit.
\par Next, consider the adiabatic expansion of the head of the jet. From Equation (11) and the scalings of the other quantities with spherical expansion in Equation (6) from \citet{mof75}, $\zeta = \omega(0.5-1.5n)$ for the jet solution. The decay from an expanding head of a conical ($\omega=1$, \citet{bla79}) Poynting jet is plotted in Figure 5 with $n=2.17$. $\zeta= 2.76$ in top panel of Figure 5 which exceeds the standard error in the least squares fit. This result is highly dependent on the value of $\omega$. The value of $\omega$ assumes that the jet head is collimated by the hoop stresses of the toroidal field, but this might not be the case in an external environment. The conclusion of Figure 5 is that the data quality is insufficient to reject either model or to claim a better fit of one model over the other.
\par The X-ray radiative losses are prodigious after day 163 (110) with an average apparent luminosity of $3.3 \times 10^{39}\rm{erg/s}$ ( $3.5 \times 10^{39}\rm{erg/s}$) until the end of year 1 post merger \citep{tro18,tro19,nyn18}. The X-ray luminosity varies in consort with the radio luminosity \citep{tro19}. Thus, the same $\delta$ might apply to both. By Equation (7) and Figure 2, the intrinsic radiative losses are one or two orders of magnitude less than the apparent luminosity. Thus, by Figure 3, the energy budget is not a concern for the jet solutions. The bottom frame of Figure 5 indicates maximum and minimum radiative losses until the end of year 1, post-merger. The minimum is just the X-ray losses with $\alpha=0.8$ from 0.3-10 keV \citep{tro19}. The maximum is more speculative and assumes a continuous powerlaw from 3 GHz to 10 KeV with $\alpha=0.584$ \citep{moo20}. The bottom frame of Figure 5 shows that the losses are too large to be consistent with the $\Gamma_{min} =5$ discrete ejection solutions. Yet, they do not exclude the $\Gamma_{min} =1$ solutions if kinetic energy is transferred to energetic particles through dissipative plasma interactions, possibly with the enveloping medium.
\begin{figure}
\begin{center}
\includegraphics[width=125 mm, angle= 0]{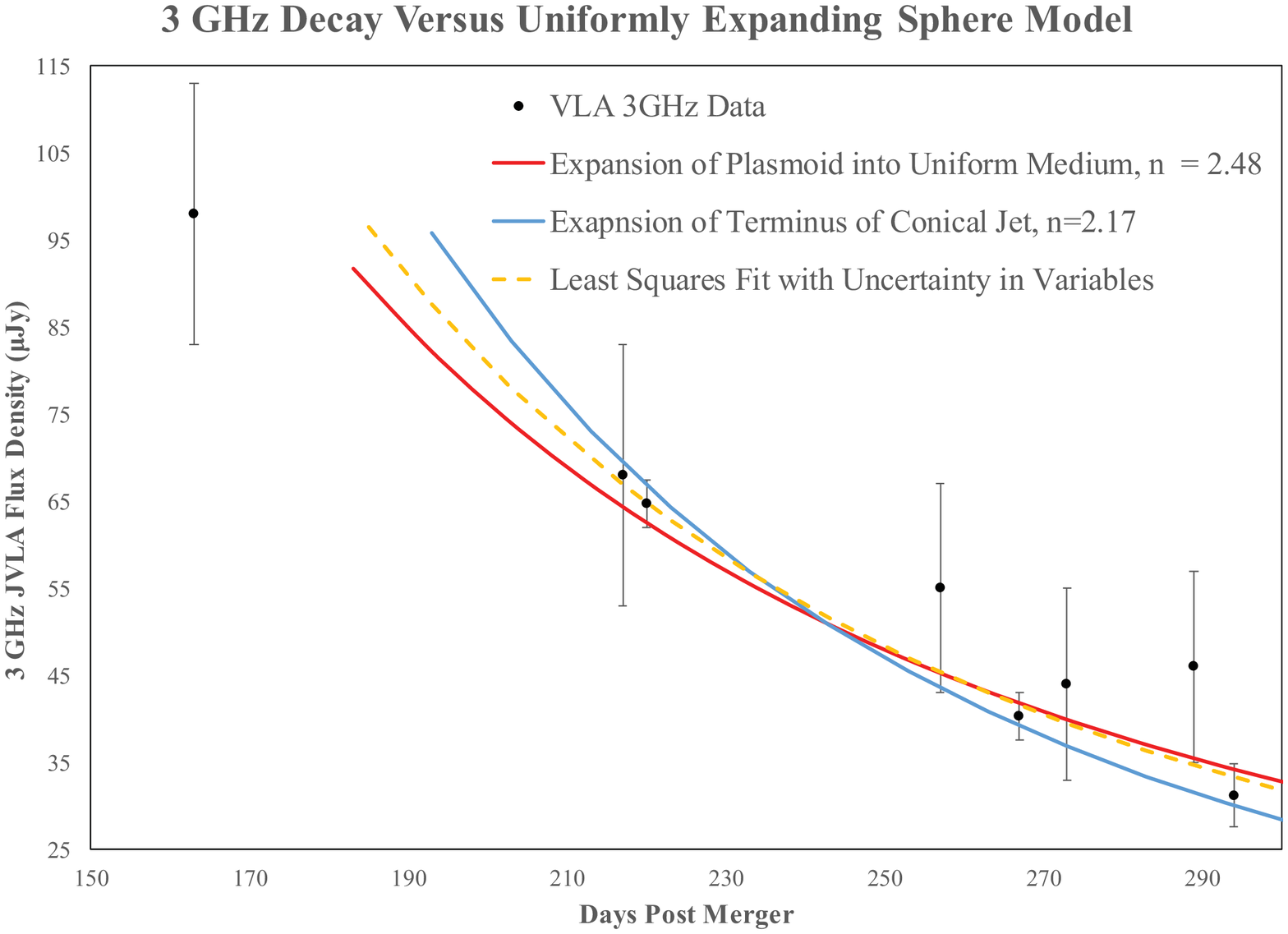}
\includegraphics[width=125 mm, angle= 0]{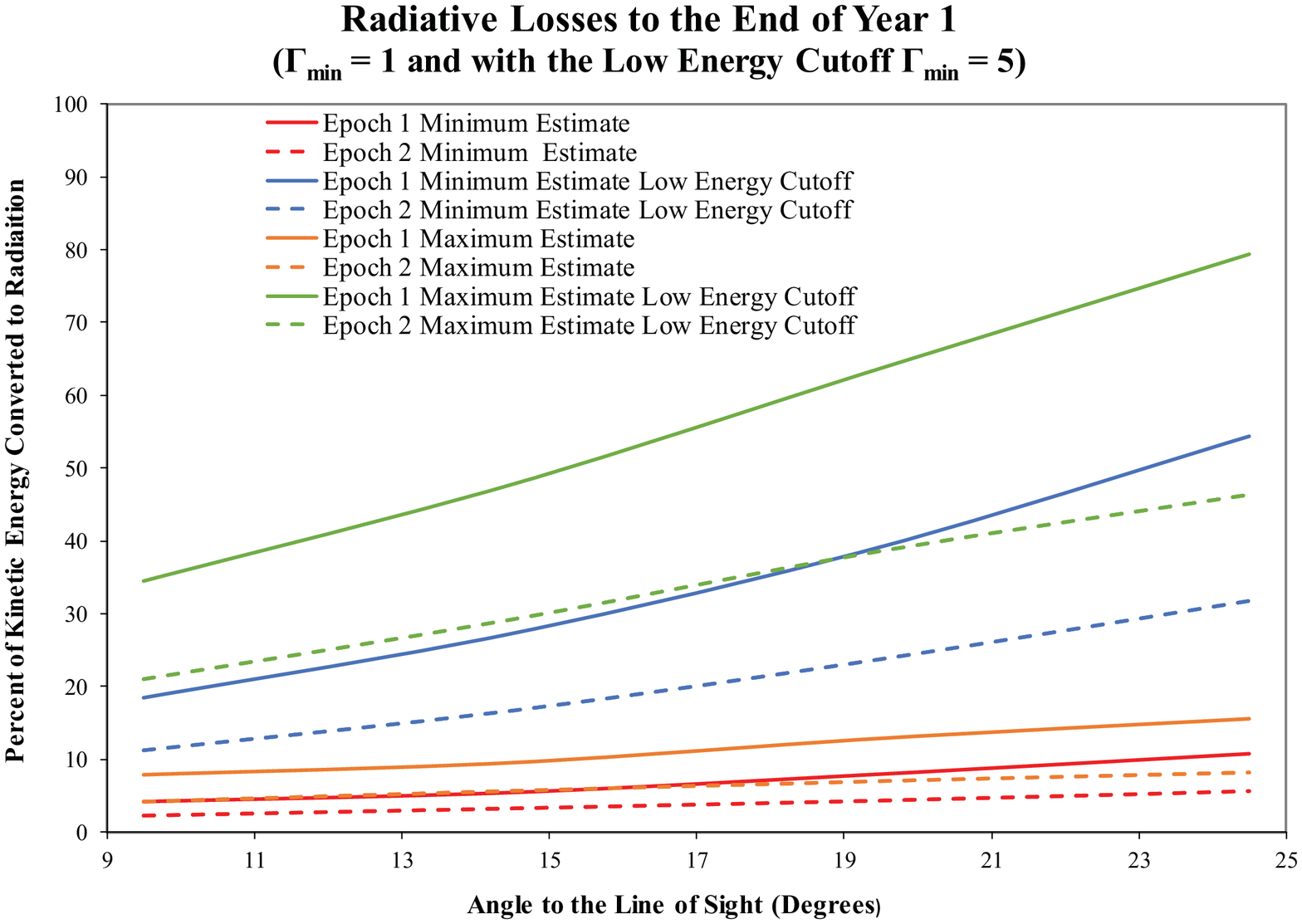}
\caption{The figure shows other constraints on the models. The top panel compares the decay from the peak of the 3 GHz light curve in the models to that of the least squares fit to the data. The bottom panel shows that radiative losses are too large for the discrete ejection model unless $\Gamma_{min}=1$. The maximum and minimum conditions are described in the text.}
\end{center}
\end{figure}
\section{Conclusion} Radio imaging data of GW170817 is used motivate a model of a discrete ejection analogous to active galactic nuclei and compact objects in the Galaxy. It is approximated as a homogeneous spherical volume moving along a preferred axis. Apparent motion in multi-epoch observations constrain the kinematics. $B$ is considered as either turbulent in a baryonic plasmoid or ordered and toroidal in a leptonic Poynting jet. Both solutions are viable based on decay light curves and energetics. The jet power in the models are $2\times 10^{39} - 8\times 10^{40} \rm{ergs/s}$. The baryonic ballistic ejection solutions have a kinetic energy of $3\times 10^{45} - 1.5\times 10^{47} \rm{ergs}$, but require $\Gamma_{min}\sim 1$ in order to support radiation losses. The main concern with the Poynting jet solutions is that they are very powerful compared to Galactic black hole (GBH) jets. The most energetic ejections from $\sim 10 M_{\odot}$ GBHs are $\sim 10^{38} \rm{ergs/s}$, if the estimates of \citet{fen99,mir94,pun12} are recomputed with the new astrometric distance and mass estimates for GRS~1915+105 \citep{rei14}. The putative post merger jet efficiency would be orders of magnitude larger than anything ever observed in the Galaxy, even for short periods ($\sim$ hours). If proton kinetic energy had been incorporated into the continuous jet models then the jet power would have been two orders of magnitude larger, making the jet power from the compact object even more extreme.

\par Neither a jet or ballistic ejection is favored by this analysis, but both types of ejections are restricted (energetically) by the models. This form of analysis provides an alternative tool to light curve time evolution. These estimates are valid only after day 110 and do not describe the ejection before this. Yet, these energy estimates can be valuable independent information for constraining the dynamics of the merger process.
\begin{acknowledgements}
I would like to thank Om Sharan Salafia, Kunal Mooley, Lekshmi Resmi and Raffaella Margutti for help with the radio data.  This work was supported by the
National Radio Astronomy Observatory, a facility of the National Science Foundation operated under cooperative agreement by
Associated Universities, Inc. Project Projects 17B-425, 18B-204,18B-302. Partial funding for this work was provided by ICRANet.
\end{acknowledgements}

\end{document}